\documentclass[showpacs,preprintnumbers,amsmath,amssymb]{revtex4}

\usepackage{epsf}
\usepackage{psfrag}
\usepackage{epsfig}
\usepackage{graphics}
\usepackage{amsmath,amssymb,graphicx}
\usepackage{slashed}

\begin{document}

\newcommand{\be}{\begin{equation}}
\newcommand{\ee}{\end{equation}}
\newcommand{\bq}{\begin{eqnarray}}
\newcommand{\eq}{\end{eqnarray}}
\newcommand{\dt}{\frac{d^3k}{(2 \pi)^3}}
\newcommand{\dtp}{\frac{d^3p}{(2 \pi)^3}}
\newcommand{\kbruto}{\hbox{$k \!\!\!{\slash}$}}
\newcommand{\pbruto}{\hbox{$p \!\!\!{\slash}$}}
\newcommand{\qbruto}{\hbox{$q \!\!\!{\slash}$}}
\newcommand{\lbruto}{\hbox{$l \!\!\!{\slash}$}}
\newcommand{\bbruto}{\hbox{$b \!\!\!{\slash}$}}
\newcommand{\parbruto}{\hbox{$\partial \!\!\!{\slash}$}}
\newcommand{\Abruto}{\hbox{$A \!\!\!{\slash}$}}
\newcommand{\Bbruto}{\hbox{$B \!\!\!{\slash}$}}

\title{\textbf{QED with chiral nonminimal coupling: aspects of the Lorentz-violating quantum corrections}}

\date{\today}

\author{A. P. Ba\^eta Scarpelli$^{(a)}$} \email[]{scarpelli.apbs@dpf.gov.br}

\affiliation{(a) Setor T\'ecnico-Cient\'{\i}fico - Departamento de Pol\'{\i}cia Federal \\
Rua Hugo D'Antola, 95 - Lapa - S\~ao Paulo - Brazil}

\begin{abstract}
An effective model for QED with the addition of a nonminimal coupling with a chiral character is investigated. This term, which
is proportional to a fixed 4-vector $b_\mu$, violates Lorentz symmetry and may originate a  CPT-even Lorentz breaking term
in the photon sector. It is shown that this Lorentz breaking CPT-even term is generated and that,in addition, the chiral nonminimal coupling requires
this term is present from the beginning. The nonrenormalizability of the model is invoked in the discussion
of this fact and the result is confronted with the one from a model with a Lorentz-violating nonminimal coupling without chirality.
\end{abstract}


\maketitle

\section{Introduction}

Lorentz violating models are being object of intensive investigation since the beginning of the nineties \cite{kostelecky1}-\cite{ferreira}.
The Standard Model Extension (SME) \cite{kostelecky1}-\cite{coleman2} proposes a wide range of possibilities in this context,
concerning both the gauge and fermion sectors. There are two particularly interesting terms in the photon sector of the SME: the CPT-odd term
of Carroll-Field-Jackiw \cite{jackiw} and a CPT-even term which is quadratic in the field strength \cite{kostelecky2}. In the context of
the SME, there have been great interest in the quantum induction of the Carroll-Field-Jackiw term \cite{CS1}-\cite{scarp1}. On the other
hand, the radiative generation of the CPT-even one (an aether-like term \cite{carroll}) has been investigated in the context of effective models which include
a Lorentz violating nonminimal (magnetic) coupling \cite{mgomes1}, \cite{mgomes2}, \cite{scarp-NM}. The perturbative generation of higher derivative
Lorentz-breaking terms from the nonminimal coupling has also been investigated \cite{petrov}.

This kind of model has been considered before in papers \cite{Belich0}-\cite{Belich3}
in the context of Relativistic Quantum Mechanics. It was used in the calculation
of corrections to the Hydrogen spectrum, from which very stringent bounds have been set up in the magnitude
of the Lorentz violating parameter \cite{Belich2}. It was also used to study the magnetic
moment generation from the nonminimal coupling, since a tiny magnetic dipole moment of elementary
neutral particles might signal Lorentz symmetry violation \cite{Belich3}.

In \cite{mgomes1} and \cite{mgomes2} the magnetic coupling has been considered in the place of the minimal coupling in a gauge violating
model. The coefficient of the induced aether-like term was shown to be regularization dependent. The inclusion of the minimal coupling
together with the magnetic one was considered in paper \cite{scarp-NM}. In this case, if a gauge invariant regularization prescription is used,
there is no quantum induction of Lorentz-violating terms in the gauge sector (CPT-even or -odd) at one-loop order. However, a gauge
violating regularization technique followed by a symmetry restoring counterterm in the Lorentz symmetric sector opens the possibility for such inductions.

In this paper, another kind of nonminimal coupling is considered, in which the background vector $b_\mu$ appears coupled to the gauge field by means
of a term of chiral character. It is shown that, in this case, there is no induction of CPT-odd terms. On the other hand, the one-loop photon
self-energy will include an aether-like CPT-even part, with a divergent coefficient. This shows that this aether-like part must be included from the beginning
in the gauge sector. This fact can be understood in terms of the nonrenormalizability of the model. If higher order contributions are considered, this term is
just the first one of an infinite list that should be included from the beginning. We discuss this fact is the context of an effective model, in which
the determination of a physical cutoff is an essential step.

This paper is organized as follows: the second section is dedicated to a general discussion on nonminimal coupling models;
the third section presents the one-loop calculation of the photon self-energy for the model with a chiral nonminimal coupling;
the conclusions are drawn in section four.

\section{Discussion on QED with nonminimal couplings}

The action of a QED model with a nonminimal coupling is written as
\be
\Sigma = \int d^4x \left\{ -\frac 14 F_{\mu \nu} F^{\mu \nu} + \bar \psi \left( i \parbruto - m
- e \Abruto - g \varepsilon^{\mu \nu \alpha \beta}
\Gamma_\beta b_\mu F_{\nu \alpha} \right) \psi \right\},
\label{action}
\ee
where $\Gamma_\beta=\gamma_\beta$ for the simple magnetic coupling and $\Gamma_\beta=\gamma_5 \gamma_\beta$ for the chiral nonminimal
coupling. Conventional QED is recovered in the limit $g \to 0$.
The action of equation (\ref{action}) describes a gauge invariant model with some interesting classical features as investigated in
\cite{Belich0}. Indeed, it is shown for the case with $\Gamma_\beta=\gamma_\beta$, that the 3-vector $\vec b$ plays the role of a kind of
magnetic dipole moment ($\vec \mu= g \vec b$). Besides, in this case there is an induction of a Aharonov-Casher (A-C) effect. On the other hand,
the chiral nonminimal coupling contributes to the interaction energy without inducting a A-C phase.

We present now a general discussion on the one-loop correction to the photon self-energy for a QED with nonminimal coupling.
First, a great simplification in the calculations is obtained if we write $B^\beta=\varepsilon^{\mu \nu \alpha \beta}
b_\mu F_{\nu \alpha}$. In the case of the simple nonminimal coupling, if the purpose is only the one-loop calculation of
the vacuum polarization tensor, it is yet possible to define the
field $\tilde A_\mu=e A_\mu+g B_\mu$, so that the new lagrangian density for the fermion sector can be written as
\be
{\cal L}_{\psi}= \bar \psi \left( i \parbruto - m
- \tilde \Abruto  \right) \psi.
\ee

In terms of $\tilde A_\mu$, the one-loop correction to the photon two-point function is identical to the one for the conventional QED. In momentum space,
we have
\be
T^{\mu\nu}(p)=\text{tr}\int_k{\gamma^\nu s(p+k)\gamma^\mu s(k)},
\label{T}
\ee
in which $s(k)$ is the fermion propagator and $\int_k$ stands for $\int d^4k/\left(2 \pi\right)^4$.
The corrections to the photon sector in the lagrangian density will be given by
\bq
&&-\frac 12 \tilde A^\mu T_{\mu \nu}(x) \tilde A^\nu \nonumber \\
&&=-\frac 12 A^\mu e^2 T_{\mu \nu}(x) A^\nu -\frac 12 A^\mu 2 eg T_{\mu \nu}(x) B^\nu
-\frac 12 B^\mu g^2 T_{\mu \nu}(x) B^\nu.
\label{magcoupling}
\eq
In this case in which chirality is absent, treated in \cite{scarp-NM}, the discussion of the induced Lorentz-violating terms
can be performed in general grounds. A general expression for $T_{\mu \nu}$ obtained by means of some regularization technique, not
necessarily gauge invariant, compatible with its Lorentz structure is
\be
T_{\mu \nu}=\left(p_\mu p_\nu -p^2 g_{\mu \nu} \right) \Pi (p^2) + \alpha m^2 g_{\mu \nu}
+ \beta p_\mu p_\nu,
\label{Tmunu}
\ee
in which $\alpha$ and $\beta$ are dimensionless constants. In eq. (\ref{magcoupling}), the first term is the traditional QED one.
The second is a Lorentz-violating CPT-odd Chern-Simons-like term, which in the on-shell limit is given by
\be
{\cal L}_{CS}=-\alpha eg m^2 \varepsilon^{\mu \nu \alpha \beta} b_\mu A_\nu F_{\alpha \beta}.
\ee
The third term in eq. (\ref{magcoupling}) is a CPT-even term, which in the on-shell limit is written as
\be
{\cal L}_{even}=-\alpha m^2 g^2 b^2 F_{\mu \nu}F^{\mu \nu}+2 \alpha m^2 g^2 \left(b_\mu F^{\mu \nu} \right)^2,
\label{Even}
\ee
where
\be
B^\mu B_\mu= 2 b^2 F_{\mu \nu}F^{\mu \nu}-4 \left(b_\mu F^{\mu \nu} \right)^2
\ee
has been obtained with the help of some properties of the L\'evi-Civit\`a tensor. We recognize in the second term of ${\cal L}_{even}$ the
Lorentz-violating aether term.

A comment is in order. The value of the constant $\alpha$ is determined by the regularization procedure used in the calculation. If a regularization technique
which preserves gauge invariance of the original QED is used, $\alpha$ will be null. However, it is always possible to choose a
gauge breaking procedure and then restore the symmetry by means of a non-symmetric counterterm. In such case, since
the Lorentz violating and Lorentz preserving parts are independent on each other, the Lorentz breaking terms would survive.
Such procedure is equivalent to use different
regularization techniques in different sectors. Nevertheless, the natural framework
is using an unique regularization in the calculation of integrals which contribute to the same amplitude. In this case, the calculation with a gauge
preserving method would not induce, at one-loop order, these two Lorentz-violating terms.

The discussion above will not apply to the case of chiral nonminimal interaction, as it will be presented in the next section.

\section{Quantum corrections to the QED with chiral nonminimal coupling}

We now turn our attention to the the one-loop quantum corrections to the photon sector in a model with a nonminimal interaction
with a chiral character. We now have the following lagrangian density for the fermion sector:
\be
{\cal L}_{\psi}= \bar \psi \left( i \parbruto - m
- e \Abruto -g \gamma_5 \Bbruto \right) \psi.
\ee
This fermion lagrangian in terms of the vector and axial-vector fields $A_\mu$ and $B_\mu$, in the case in which the
fields do not depend on each other,
has been vastly investigated (see, for example, \cite{Bardeen} and \cite{Bonora}). The one-loop corrections to the photon sector will be given by
\be
-\frac 12 A^\mu e^2 T^{VV}_{\mu \nu}(x) A^\nu -\frac 12 A^\mu 2 eg T^{AV}_{\mu \nu}(x) B^\nu
-\frac 12 B^\mu g^2 T^{AA}_{\mu \nu}(x) B^\nu,
\label{chicoupling}
\ee
where the upper indices $A$ and $V$ refer to the axial and vectorial vertices, so that $T^{VV}_{\mu \nu}=T_{\mu \nu}$. In momentum space, we have
\be
T^{AV}_{\mu \nu}(p)=\int_k{\text{tr}\left\{\gamma^\nu s(p+k)\gamma_5\gamma^\mu s(k)\right\}}
\label{TAV}
\ee
and
\be
T^{AA}_{\mu \nu}(p)=\int_k{\text{tr}\left\{\gamma_5\gamma^\nu s(p+k)\gamma_5\gamma^\mu s(k)\right\}}.
\ee
The second term in (\ref{chicoupling}), which would be CPT-odd, is actually identically null, since
in eq. (\ref{TAV}), we have in the numerator
\bq
&&\text{tr}\left\{\gamma_5\gamma^\rho (\pbruto + \kbruto +m)\gamma^\mu (\kbruto+m)\right\}=
\text{tr}\left\{\gamma_5\gamma^\rho (\pbruto + \kbruto)\gamma^\mu \kbruto\right\} + \nonumber \\
&&+m^2 \text{tr}\left\{\gamma_5\gamma^\rho \gamma^\mu \right\}
+ m\text{tr}\left\{\gamma_5\gamma^\rho (\pbruto + \kbruto)\gamma^\mu \right\}
+m\text{tr}\left\{\gamma_5\gamma^\rho \gamma^\mu \kbruto\right\} \nonumber \\
&&=4i \varepsilon^{\rho \kappa \mu \lambda}(p+k)_\kappa k_\lambda =
4i \varepsilon^{\rho \kappa \mu \lambda}p_\kappa k_\lambda.
\eq
Under integration, this will vanish due to the antisymmetry of the L\'evi-Civit\`a tensor.

We are left with the CPT-even term of (\ref{chicoupling}). First, we can write
\be
\text{tr}\left\{\gamma_5\gamma^\delta (\pbruto+\kbruto+m)\gamma_5\gamma^\rho (\kbruto+m)\right\}
= \text{tr}\left\{\gamma^\delta (\pbruto+\kbruto+m)\gamma^\rho (\kbruto+m)\right\} -8m^2 g^{\rho \delta},
\ee
so that
\be
T^{AA}_{\mu \nu}=T_{\mu \nu}-8m^2 g_{\mu \nu} I
\label{aa}
\ee
where
\be
I= \int^\Lambda \frac{d^4k}{(2 \pi)^4} \frac {1}{(k^2-m^2)\left[(p-k)^2-m^2\right]}
\label{I}
\ee
is a divergent integral and $\Lambda$ is to indicate that some regularization prescription is applied. We have to note that
the contribution of the first term in (\ref{aa}) is identical to the one calculated with the nonminimal coupling without chirality.
The divergent integral $I$ can be evaluated by any regularization method. In order to make the regularization independence manifest,
we may write equation (\ref{I}) in a way that divergences are expressed in terms of the loop momentum only, as in
Implicit Regularization \cite{papersIR}:
\be
I= I_{log}\left(\lambda^2\right)-b Z_0(p^2,m^2,\lambda^2),
\ee
where
\be
I_{log}\left(\lambda^2\right)=\int^\Lambda \frac{d^4k}{(2 \pi)^4} \frac {1}{(k^2-\lambda^2)^2},
\ee
\be
Z_0(p^2,m^2,\lambda^2)=\int_0^1 dx\,\ln{\left(\frac{p^2x(1-x)-m^2}{\left(-\lambda^2\right)}\right)},
\ee
$b=i/(4 \pi)^2$ and $\lambda^2$ is an arbitrary ultraviolet mass scale. Since we are interested in the on-shell limit, we will be left with
\be
T^{AA}_{\mu \nu}=m^2 g_{\mu \nu}
\left[-8 I_{log}\left(\lambda^2\right) +\alpha +8b \ln{\left(\frac{m^2}{\lambda^2}\right)}\right] \equiv F\left(m^2, \lambda^2\right)g_{\mu \nu},
\ee
where $\alpha$ is defined in equation (\ref{Tmunu}). So, the CPT-even term will be given by
\bq
&&{\cal L}_{even}=-\frac 12 F\left(m^2, \lambda^2\right) B_\mu B^\mu   \nonumber \\
&&= -F\left(m^2, \lambda^2\right)\left[b^2 F_{\mu \nu}F^{\mu \nu}-2 \left(b_\mu F^{\mu \nu} \right)^2 \right].
\label{result}
\eq
The Lorentz-violating second  term above can be mapped in the CPT-even term proposed in \cite{kostelecky2},
\be
{\cal L}_{even}=-\frac 14 \kappa_{\mu \nu \alpha \beta} F^{\mu \nu} F^{\alpha \beta},
\ee
as long as we establish the relation
\be
\kappa_{\mu \nu \alpha \beta}= -2 F\left(m^2, \lambda^2\right) \left(
g_{\mu \alpha} b_\nu b_\beta - g_{\nu \alpha} b_\mu b_\beta
+ g_{\nu \beta} b_\mu b_\alpha - g_{\mu \beta} b_\nu b_\alpha
 \right).
\ee

We are now in position to discuss the result of equation (\ref{result}). First, as discussed in the last section, the constant $\alpha$
vanishes if a gauge invariant regularization technique is used, although the gauge symmetry of the model could be preserved if the Lorentz
invariant and Lorentz-violating sectors are treated independently. However, the value of $\alpha$ is irrelevant here, since it can be absorbed
in the other finite term, which depend on an arbitrary mass scale parameter.

Second, the presence of a divergent term in the CPT-even coefficient is an important point to be analyzed. This indicates that the original
classical action must contain such a term. In other words,
the inclusion of a chiral nonminimal coupling in a modified QED requires the presence of the aether term from the beginning.
This divergent factor multiplies also a Maxwell term. This means that the Lorentz preserving sector is
also affected by the presence of the nonminimal interaction of chiral character. This is in contrast with the case of standard nonminimal coupling treated
in papers \cite{mgomes1}, \cite{mgomes2} and \cite{scarp-NM}. In that case, the correction to the Maxwell term (and also the induced aether term)
is finite and arbitrary, with the possibility of being set to zero.

Last but not least, we must take into account that our model is nonrenormalizable. We have carried out an one-loop calculation and, at this order
in the perturbative expansion, it has been shown that a new term which violates Lorentz symmetry and is CPT-even should be included in the classical
action. If we go beyond the one-loop order, certainly new other terms will have to be considered. The nonrenormalizability of the model
tells us that there is no a finite number of counterterms that will be sufficient to renormalize the theory. So, if we would like to
deal with this effective model, we will have to stop at one-loop order. For this, it is necessary to find a cutoff energy $\Lambda$. This can be done
like in the case of the simple nonminimal coupling, discussed bellow.

We have to note that higher order terms in the coupling constant will allow for higher power contributions in the Lorentz violating parameter, 
with an increasing of the degree of divergence of the integrals. However, as demonstrated in \cite{Belich2} for the vectorial nonminimal coupling, 
the magnitude of the background vector is extremely small. We can impose the reasonable condition for the recovering of QED, $|b^2| \Lambda^2 << 1$. 
Since the effect of the divergences can be seen in a simplified form by substituting $m^2$ by $\Lambda^2$ in the
coefficients, higher order calculations will furnish us higher powers of $|b^2| \Lambda^2$.
So, although we can not prevent the proliferation of new terms beyond one-loop order,
the predictability of such effective model is assured by the cutoff inequality
above. This happens because at the same loop order each nonminimal vertex contributes with
one factor of $b_\mu$, whereas higher loop orders with a fixed number of nonminimal vertices are controlled by the smallness of the coupling constant.

In an effective model, the cutoff energy is an important parameter which should be established on physical grounds.
Important features of a Quantum Field Theory, like causality and stability, can be lost at high energies \cite{causality1},
\cite{causality2}. The condition imposed by the inequality $|b^2| \Lambda^2 << 1$ is so that higher power terms in $b_\mu$
become less significant. In \cite{Belich2}, it has been established a bound such that $g \cdot |b_\mu|<10^{-32}\,(eV)^{-1}$
for the simple magnetic coupling.
This bound and the inequality we propose above assure that such effective model is not considered at energies
beyond the Planck scale.

So, it is desirable that a calculation such that of \cite{Belich2} would be performed in order to establish a bound for the Lorentz
violating parameter in the case of a chiral nonminimal coupling. In this case, we could consider the one-loop calculation meaningful.

\section{Concluding Comments}

An effective model for QED with the addition of a chiral nonminimal coupling has been investigated. This term, which
is proportional to a fixed 4-vector $b_\mu$, violates Lorentz symmetry and originates a  CPT-even Lorentz breaking term
in the photon sector. Besides, since the coefficient of this quantum correction is divergent, such a model
requires the presence of this aether term from the beginning in the classical action.
This divergent factor multiplies also a Maxwell term. This means that the Lorentz preserving sector is
also affected by the presence of the nonminimal interaction of chiral character. This is in contrast with the case of standard nonminimal coupling treated
in the papers \cite{mgomes1}, \cite{mgomes2} and \cite{scarp-NM}. In that case, the correction to the Maxwell term is finite and arbitrary, with
the possibility of being set to zero.

The fact that the aether term should be present in the classical action from the beginning is problematic. If we go beyond
one-loop order, the nonrenormalizability of the model
tells us that there is no a finite number of counterterms that will be sufficient to renormalize the theory. So, if we would like to
deal with this effective model, we will have to stop at one-loop order. For this, it is necessary to find a cutoff energy $\Lambda$. This can be done
like in the case of the simple nonminimal coupling. In an effective model, the cutoff energy is an important parameter which should be established on physical grounds.Important features of a Quantum Field Theory, like causality and stability, can be lost at high energies \cite{causality1},
\cite{causality2}. The condition imposed by the inequality $|b^2| \Lambda^2 << 1$ is so that higher power terms in $b_\mu$
become less significant. In \cite{Belich2}, it has been established a bound such that $g \cdot |b_\mu|<10^{-32}\,(eV)^{-1}$
for the simple magnetic coupling.
This bound and the inequality we propose above assure that such effective model is not considered at energies
beyond the Planck scale.

So, it is desirable that a calculation such that of \cite{Belich2} would be performed in order to establish a bound for the Lorentz
violating parameter in the case of a chiral nonminimal coupling. In this case, we could consider the one-loop calculation meaningful.

\subsection*{Acknowledgements}

This work was partially supported by CNPq. The author wish to thank Marcos Sampaio and Prof. J. A. Helayel-Neto
for illuminating discussions.



\begin{thebibliography}{99}
\bibitem{kostelecky1}  D. Colladay and V. A. Kostelecky, \textit{{Phys. Rev.
\textbf{D}}} \textbf{55}, (1997) 6760.

\bibitem{kostelecky2}  D. Colladay and V. A. Kostelecky, \textit{{Phys. Rev.
\textbf{D}}} \textbf{58}, (1998) 116002.

\bibitem{coleman1}  S. Coleman and S. L. Glashow, \textit{{Phys. Lett. \textbf{B}}}
\textbf{405}, (1997) 249.

\bibitem{coleman2} S. Coleman and S. L. Glashow, \textit{{Phys. Rev.
\textbf{D}}} \textbf{59}, (1999) 116008.

\bibitem{carroll} S. Carroll and H. Tam, \textit{{Phys. Rev. \textbf{D}}} \textbf{78}, (2008) 044047.

\bibitem{mgomes1} M. Gomes, J. R. Nascimento, A. Yu. Petrov, A. J. da Silva,
\textit{{Phys. Rev. \textbf{D}}} \textbf{81}, (2010) 045018.

\bibitem{mgomes2} M. Gomes, J.R. Nascimento, A.Yu. Petrov, A.J. da Silva,
``On the aether-like Lorentz-breaking action for the electromagnetic field'' [arxiv:1008.0607].

\bibitem{scarp-NM} G. Gazzola, H.G. Fargnoli, A.P. Ba\^eta Scarpelli, Marcos Sampaio, M.C. Nemes,
\textit{{J. Phys. \textbf{G}}} \textbf{39} (2012) 035002 [arXiv:1012.3291].

\bibitem{petrov} T. Mariz, J.R. Nascimento, A.Yu. Petrov, ``On the perturbative generation of the higher-derivative Lorentz-breaking terms''
[arXiv:1111.0198].

\bibitem{Belich0} H. Belich, T. Costa-Soares, M. M. Ferreira Jr., J.A. Helayel-Neto,
\textit{{Eur. Phys. J. \textbf{C}}}
\textbf{41}, (2005) 421 [hep-th/0410104].

\bibitem{Belich1} H. Belich, T. Costa-Soares, M. M. Ferreira Jr., J.A. Helayel-Neto,
\textit{{Eur. Phys. J. \textbf{C}}}
\textbf{42}, (2005) 127 [hep-th/0411151].

\bibitem{Belich2} H. Belich, T. Costa-Soares, M. M. Ferreira Jr., J.A. Helayel-Neto, F. M. O Moucherek,
\textit{{Phys. Rev. \textbf{D}}} \textbf{74}, (2006) 065009 [hep-th/0604149].

\bibitem{Belich3} H. Belich, L.P. Colatto, T. Costa-Soares, J.A. Helayel-Neto, M.T.D. Orlando,
\textit{{Eur. Phys. J. \textbf{C}}}
\textbf{62}, (2009) 425 [arXiv:0806.1253].

\bibitem{5} B. Goncalves, Y. N. Obukhov, I. L. Shapiro, \textit{{Phys.Rev. \textbf{D}}} \textbf{80}, (2009) 125034.

\bibitem{causality1} V. A. Kostelecky and R. Lehnert, \textit{{Phys. Rev. \textbf{D}}} \textbf{63}, (2001) 065008.

\bibitem{causality2} J. Alfaro, A. A. Andrianov, M. Cambiaso, P. Giacconi, R. Soldati, \textit{{Int.J.Mod.Phys. \textbf{A}}}
\textbf{25}, (2010) 3271;

\bibitem{jackiw} S.M. Carroll, G.B. Field and R. Jackiw, \textit{{Phys. Rev. \textbf{D}}} \textbf{41}, (1990) 1231.

\bibitem{8} A. P. Ba\^eta Scarpelli, H. Belich, J. L. Boldo, J. A. Helayel-Neto,
\textit{{Phys. Rev. \textbf{D}}} \textbf{67}, (2003) 085021;
A. P. Ba\^eta Scarpelli and J. A. Helayel-Neto, \textit{{Phys.Rev. \textbf{D}}} \textbf{73}, (2006) 105020.

\bibitem{9} C. Adam and F. R. Klinkhamer, \textit{{Nucl. Phys. \textbf{B}}} \textbf{607}, (2001) 247;
C. Adam and F. R. Klinkhamer, \textit{{Nucl. Phys. \textbf{B}}} \textbf{657}, (2003) 214.

\bibitem{10} A.A. Andrianov and R. Soldati, \textit{{Phys. Rev. \textbf{D}}} \textbf{51}, (1995) 5961;
A. A. Andrianov and R. Soldati, \textit{{Phys. Lett. \textbf{B}}} \textbf{435}, (1998) 449;
A. A. Andrianov, R. Soldati and L. Sorbo, \textit{{Phys. Rev. \textbf{D}}} \textbf{59}, (1998) 025002;
A. A. Andrianov, D. Espriu, P. Giacconi, R. Soldati, \textit{{J. High Energy Phys.}}
\textbf{0909}, (2009) 057.


\bibitem{12} M.S. Berger, V. A. Kostelecky, \textit{{Phys.Rev. \textbf{D}}} \textbf{65}, (2002) 091701;
H. Belich , J.L. Boldo, L.P. Colatto, J.A. Helayel-Neto,
A.L.M.A. Nogueira, \textit{{Phys.Rev. \textbf{D}}} \textbf{68}, (2003) 065030.

\bibitem{CS1} R. Jackiw and V. A. Kostelecky, \textit{{Phys. Rev. Lett.}} \textbf{82}, (1999) 3572.

\bibitem{CS2} J. M. Chung and B. K. Chung, \textit{{Phys. Rev. \textbf{D}}} \textbf{63}, (2001) 105015.

\bibitem{CS3} J.M. Chung, \textit{{Phys.Rev. \textbf{D}}} \textbf{60}, (1999) 127901.

\bibitem{CS4} W. F. Chen, \textit{{Phys. Rev. \textbf{D}}} \textbf{60}, (1999) 085007.

\bibitem{CS5} M. Perez-Victoria, \textit{{Phys. Rev. Lett.}} \textbf{83}, (1999) 2518.

\bibitem{CS6} M. Perez-Victoria, \textit{{J. High. Energy Phys.}} \textbf{0104}, (2001) 032.

\bibitem{CS7}O.A. Battistel and G. Dallabona, \textit{{Nucl. Phys. \textbf{B}}} \textbf{610}, (2001) 316;
\textit{{J. Phys. \textbf{G}}} \textbf{27}, (2001) L53;
O.A. Battistel and G. Dallabona, \textit{{J. Phys. \textbf{G}}} \textbf{28}, (2002) L23.

\bibitem{CS8} A. P. Ba\^eta Scarpelli, M. Sampaio, M. C. Nemes, and B. Hiller,
\textit{{Phys. Rev. \textbf{D}}} \textbf{64}, (2001) 046013.

\bibitem{CS9} T. Mariz, J.R. Nascimento, E. Passos, R.F. Ribeiro and F.A. Brito,
\textit{{J. High. Energy Phys.}} \textbf{0510} (2005) 019.

\bibitem{CS10} J. R. Nascimento, E. Passos, A. Yu. Petrov, F. A. Brito,
\textit{{J. High. Energy Phys.}} \textbf{0706}, (2007) 016.

\bibitem{CS11} B. Altschul, \textit{{Phys. Rev. \textbf{D}}} \textbf{70}, (2004) 101701.

\bibitem{CS12} F.A. Brito, J.R. Nascimento, E. Passos, A.Yu. Petrov,
\textit{{Phys. Lett. \textbf{B}}} \textbf{664}, (2008) 112.

\bibitem{CS13} Oswaldo M. Del Cima, J. M. Fonseca, D.H.T. Franco, O. Piguet,
\textit{{Phys. Lett. \textbf{B}}} \textbf{688}, (2010) 258.

\bibitem{CS14} F.A. Brito, L.S. Grigorio, M.S. Guimaraes, E. Passos, C. Wotzasek,
\textit{{Phys.Rev. \textbf{D}}} \textbf{78}, (2008) 125023.

\bibitem{scarp1} A.P. Ba\^eta Scarpelli, M. Sampaio, M.C. Nemes, B. Hiller,
\textit{{Eur. Phys. J. \textbf{C}}} \textbf{56}, (2008) 571.

\bibitem{14} R. Lehnert and R. Potting, \textit{{Phys. Rev. Lett.}} \textbf{93}, (2004) 110402.

\bibitem{18} R. Casana, M. M. Ferreira Jr. and J. S. Rodrigues,
\textit{{Phys. Rev. \textbf{D}}} \textbf{78}, (2008) 125013;
F.A. Brito, L.S. Grigorio, M.S. Guimaraes, E. Passos, C. Wotzasek,
\textit{{Phys. Lett. \textbf{B}}} \textbf{681}, (2009) 495.

\bibitem{20} V. A. Kostelecky and M. Mewes, \textit{{Phys. Rev. Lett.}} \textbf{87}, (2001) 251304.

\bibitem{21} V. A. Kostelecky and M. Mewes, \textit{{Phys. Rev. \textbf{D}}} \textbf{66}, (2002) 056005.

\bibitem{22} V. A. Kostelecky and M. Mewes, \textit{{Phys. Rev. \textbf{D}}} \textbf{80}, (2009) 015020.

\bibitem{23} V. A. Kostelecky and M. Mewes, \textit{{Phys. Rev. Lett.}} \textbf{97}, (2006) 140401.

\bibitem{25} V.A. Kostelecky and R. Potting, \textit{{Phys.Rev. \textbf{D}}} \textbf{79}, (2009) 065018;
Q. G. Bailey, \textit{{Phys.Rev. \textbf{D}}} \textbf{82}, (2010) 065012;
V.B. Bezerra, C.N. Ferreira, J.A. Helayel-Neto, \textit{{Phys.Rev. \textbf{D}}} \textbf{71}, (2005) 044018.

\bibitem{26} T. Jacobson, S. Liberati, D.Mattingly, \textit{{Nature}} \textbf{424}, (2003) 1019;
T. Jacobson, S. Liberati, D.Mattingly, \textit{{Phys.Rev. \textbf{D}}} \textbf{67}, (2003) 124011.

\bibitem{32} C.D. Carone, M. Sher, and M. Vanderhaeghen,
\textit{{Phys. Rev. \textbf{D}}} \textbf{74}, (2006) 077901.

\bibitem{34} J.-P. Bocquet et at., \textit{{Phys. Rev. Lett.}} \textbf{104}, (2010) 241601.

\bibitem{37} R. Casana, M.M. Ferreira Jr, A. R. Gomes, P. R. D. Pinheiro,
\textit{{Phys. Rev. \textbf{D}}} \textbf{80}, (2009) 125040.

\bibitem{38} B. Altschul, \textit{{Phys. Rev. Lett.}} \textbf{98}, (2007) 041603.

\bibitem{39} R. Casana, M.M. Ferreira Jr, Carlos E. H. Santos,
\textit{{Phys. Rev. \textbf{D}}} \textbf{78}, (2008) 105014.

\bibitem{ferreira} Rodolfo Casana, Manoel M. Ferreira, Adalto R. Gomes, Frederico E. P. dos Santos,
\textit{{Phys. Rev. \textbf{D}}} \textbf{82}, (2010) 125006 [arxiv:1010.2776].

\bibitem{jackiw2} R. Jackiw, \textit{{Int. J. Mod. Phys.\textbf{B}}} \textbf{14}, (2000) 2011 [hep-th/9903044].

\bibitem{Bardeen} W. A. Bardeen, \textit{{Phys. Rev.}} \textbf{184}, (1969) 1848.

\bibitem{Bonora} A. Andrianov, L. Bonora, \textit{{Nucl. Phys.\textbf{B}}} \textbf{233}, (1984) 232.

\bibitem{papersIR}
A. P. Ba\^{e}ta Scarpelli, M. Sampaio and M. C. Nemes,
\textit{{Phys. Rev. \textbf{D}}} \textbf{63}, (2001) 046004;
E. W. Dias, A. P. Ba\^eta Scarpelli, L. C. T. Brito, Marcos Sampaio, M. C. Nemes,
\textit{{Eur. Phys. J. \textbf{C}}} \textbf{55}, (2008) 667;
E. W. Dias, A. P. Ba\^eta Scarpelli, L. C. T. Brito, H. G. Fargnoli,
\textit{{Braz. J. Phys.}} \textbf{40}, (2010) 228;
A. L. Cherchiglia, Marcos Sampaio, M. C. Nemes,
\textit{{Int. J. Mod. Phys. \textbf{A}}} \textbf{26}, (2011) 2591;
O. A. Battistel, {\it PhD thesis}, Federal University of
Minas Gerais (2000).

\end{thebibliography}
\end{document}